\documentclass[reprint, nofootinbib, amsmath,amssymb, aps, prd]{revtex4-2}
\usepackage[dvipsnames,table,xcdraw]{xcolor}
\usepackage[colorlinks,urlcolor=blue,citecolor=blue,linkcolor=blue, pdfborder={0 0 0}]{hyperref}
\usepackage{graphicx}
\usepackage{url}
\usepackage{bm}
\usepackage{comment}
\usepackage{float}
\begin{document}

% \preprint{APS/zQED}

\title{Scalar Quasi-Normal Modes in Black Hole Gravitational Lensing}
\author{Chengjiang Yin$^{1,2}$}
\thanks{These authors contributed equally to this paper.}
\author{Zihao Lin$^{1,2}$}
\thanks{These authors contributed equally to this paper.}
\author{Jian-hua He$^{1,2}$}
\thanks{Corresponding author: \href{mailto:hejianhua@nju.edu.cn}{hejianhua@nju.edu.cn}}
\affiliation{$^1$School of Astronomy and Space Science, Nanjing University, Nanjing 210093, P. R. China}
\affiliation{$^2$Key Laboratory of Modern Astronomy and Astrophysics (Nanjing University), Ministry of Education, Nanjing 210023, P. R. China}

\date{\today}

\begin{abstract}
We investigate the excitation of quasi-normal modes (QNMs) in gravitational lensing by a Schwarzschild black hole using a scalar field model. By employing a time-domain mode-sum method, we analyze the complex interplay between an incident burst signal and the black hole spacetime. We find that the incident waves can non-resonantly excite a substantial number of high-$l$ modes, with amplitudes for modes as high as $l=20$ remaining significant compared to the fundamental $l=0$ mode. We confirm through QNM template fitting that  the late-time behaviors of these excited modes are indeed QNMs. After passing through the black hole, we find that the lensed waves form a highly directional and coherent Gaussian beam whose cross-sectional intensity profile is well-described by a Gaussian profile. Unlike spherical waves, this beam's amplitude does not decrease with distance from the black hole but remains nearly constant in the near-field region. Moreover, due to the superposition of numerous QNMs, oscillations largely cancel each other out. The lensed temporal waves do not exhibit typical oscillatory patterns. 
\end{abstract}

\maketitle

\section{Introduction} 
When a black hole (BH) is disturbed, it does not merely passively absorb the perturbation. Instead, the BH itself ``rings", radiating away excess energy as gravitational waves. These characteristic oscillations are known as quasi-normal modes (QNMs)  (see reviews and references therein~\cite{Hans-Peter_Nollert_1999,Kokkotas:1999bd,Berti_2009,RevModPhys.83.793, Berti:2025hly}). QNMs are expected to occur in two primary astrophysical scenarios: (1) the ``ringdown" stage following a binary BH merger, and (2) the accretion of substantial clumps of matter onto a BH~\cite{PhysRevD.5.2932,FERRARI1981381}.

Besides the ringdown stage and the accretion of matter, QNMs can also be excited by gravitational waves (GWs) passing through a BH. Similar to light, GWs undergo significant trajectory and waveform alterations due to the BH's gravity~\cite{He2022PRD,Yin2024PRL}. Although this phenomenon involves complex wave dynamics, recent celebrated mathematical proofs have demonstrated a general behavior for waves arising from regular initial metric perturbations in Schwarzschild spacetime: their energy remains pointwise bounded at all times and decays according to an inverse polynomial law with respect to time~\cite{Dafermos:2016uzj}(The analogous result for scalar fields was proven much earlier by Kay and Wald~\cite{Kay_Wald_1987}). Consequently, due to this bounded and decaying behavior, GWs around a BH are unlikely to trigger significant nonlinear effects~\cite{Dafermos:2021cbw}, meaning their behavior can be well-described by linear theory. This inherent linearity potentially provides a cleaner avenue for testing linear perturbation theory compared to the ringdown stage, where nonlinear dynamics can introduce significant complexities and potentially challenge the validity of tests based on linear theory~\cite{NonlinearRingdown1,NonlinearRingdown2}.

Lensed GWs has primarily been studied in two regimes: the weak-field limit and BH spacetimes. In the weak-field limit, the propagation of GWs is typically investigated using frequency-domain wave optics~\cite{Suyama:2005mx,PhysRevLett.80.1138,PhysRevD.98.083005,Ruffa_1999,DePaolis:2002tw,Zakharov_2002,Liao:2019aqq,Macquart:2004sh,Dai:2018enj,PhysRevD.90.062003,Christian,Meena:2019ate,2021PhRvD.103j4055W,Hongsheng:2018ibg,PhysRevD.111.103539,PhysRevD.111.103539,PhysRevD.108.043527}.
In contrast, investigations in BH spacetimes primarily rely on frequency-domain scattering theory. However, this latter approach faces a significant challenge: the conventional partial-wave method suffers from a long-standing divergence problem, wherein the partial-wave series fails to converge~\cite{marquez_divergence_1972}. While several workarounds have been proposed, they are subject to some limitations. 
For instance, regularization schemes like Cesàro summation break down for forward scattering $\theta=0$.
As a result, current analyses based on
scattering theory are restricted to off-axis cases~\cite{
PhysRevD.110.044054,
6h6r-46cd}.
Alternatively, one can truncate the partial-wave series in the eikonal limit~\cite{Nambu:2015aea}, but the resulting closed form expression is only valid for high-$l$ modes.
A third method involves positioning the source and observer at a finite distance~\cite{Kanai:2013rga,PhysRevD.110.124011,Pijnenburg:2022pug,xsxj-9vdw}, which ensures convergence because even a divergent wave function remains finite at finite distances. However, the standard relationship between the observed GW waveform and the Weyl scalar $\psi_4$ (or Regge-Wheeler-Zerilli-Moncrief variables~\cite{Regge_Wheeler,Zerillia,Zerillib,MONCRIEF1974343,1973ApJ...185..635T,Chandrasekhar:1985kt})
\begin{align}
\psi_4=\ddot{h}_{+}-i \ddot{h}_{\times}\,,
\end{align}
is valid only at null infinity relative to the BH~\cite{PhysRevD.82.104057}. 
Therefore, it is necessary to extrapolate the results at a finite distance out to infinity. 
One common approach is direct extrapolation~\cite{PhysRevD.82.104057}.
Another method involves compactification, which includes null infinity within the computational domain~\cite{PhysRevX.1.021017}. However, whether these methods can be applied consistently in the context of BH lensing without encountering divergence issues at infinity requires further investigation.

In contrast, rigorous mathematical analyses \cite{Kay_Wald_1987,Dafermos:2016uzj}  suggest the time-domain is free from these divergent problems. Nevertheless, even within the framework of linear theory, waves propagating in the curved spacetime of strong gravity display intricate behaviors~\cite{He2022PRD}. Besides the excitation of QNMs, the spacetime curvature significantly influences the trajectory of incoming waves. Moreover, due to the failure of Huygens' principle in BH spacetimes, propagating finite wave packets do not form sharp trailing wavefronts; instead, they develop continuously decaying tails (see, e.g., the detailed discussions in~\cite{Friedlander:2010eqa}). As a result, gravitational wave signals lensed by a BH are a superposition of three effects: trajectory bending, QNM excitation, and tail formation. This intricate interplay can be further complicated by complex three-dimensional (3D) spatial geometry of the lensing system. 

In this work, we investigate this complex behavior, particularly the QNM signals in BH gravitational lensing, using the time-domain mode-sum method, with results cross-validated by high-resolution 3D numerical simulations. As a first step, we conduct numerical experiments with a scalar field model 
\begin{equation}
   \Box h = (-g)^{-1/2} \partial_\mu \left[(-g)^{1/2} g^{\mu\nu}\partial_\nu \right]h= 0\,,
\label{eq:scalarfield}
\end{equation}
where $h$ is a scalar field and $g$ denotes the determinant of the metric $g_{\mu\nu}$. This model was chosen because, compared to gravitational wave tensor field equations, it allows for substantially higher resolutions with limited computational resources, thereby enhancing the reliability of our numerical results. Furthermore, it greatly simplifies the construction of initial conditions for complex, non-spherical geometries. Throughout this paper, we adopt the geometric unit $c=G=1$, in which $1 M_{\odot}=4.92535\times 10^{-6} {\rm Sec}\,$.

\section{Time-domain mode-sum method}  We consider solving Eq.(\ref{eq:scalarfield}) in the Schwarzschild spacetime under isotropic coordinate
\begin{equation}
ds^2 = -\left(\frac{1 - \frac{M}{2\rho}}{1 + \frac{M}{2\rho}}\right)^2 dt^2 
+ \left(1 + \frac{M}{2\rho}\right)^4 
\left( dx^2 + dy^2 + dz^2 \right),
\end{equation}
where $\rho=\sqrt{x^2+y^2+z^2}$ is the radius. Since the spatial part of the background metric is spherically symmetric, it is convenient to decompose $h$ into spherical harmonics
\begin{equation}
   h(t,\rho, \theta, \phi) =\frac{1}{r(\rho)} \sum_{l=0}^{\infty} \sum_{m=-l}^{l} u_{l m}(t,\rho) Y_{l}^{m}(\theta, \phi)
\label{eq:Spherical_Harmonic}\,.
\end{equation}
The pre-factor $1/r(\rho)$, with $r(\rho)=(M+2\rho)^2/(4\rho)$ being the radius in Schwarzschild coordinates, is included to ensure the definition of  $u_{lm}$ is consistent with the literature~\cite{Hans-Peter_Nollert_1999,Berti_2009,Kokkotas:1999bd,RevModPhys.83.793,Berti:2025hly}. In this paper, we only consider azimuthally symmetric waves, for which only modes with $m=0$ contribute to the decomposition. Therefore, we denote $u_{lm}$ simply as $u_l$ hereafter. The radial equation is then given by
\begin{align}
\frac{\partial^2 u_l}{\partial t^2} &= \tilde{c}^2  \frac{\partial^2 u_l}{\partial \rho^2} +  \tilde{c}^2\left[  \frac{2M(4\rho-M)}{\rho(4\rho^2 - M^2)} \right] \frac{\partial u_l}{\partial \rho} \nonumber\\ &- \tilde{c}^2\left[ \frac{l(l+1)}{\rho^2}+ \frac{8\rho M}{\rho^2(2\rho+M)^2}  \right] u_l 
\label{eq:timedomain_wave}\,,
\end{align}
where $\tilde{c} = 4\rho^2 (2\rho-M)/(2\rho+M)^3$ is the speed of light in isotropic coordinate~\cite{He2022PRD}.
\subsection{Initial and boundary conditions} 
To calculate the 3D propagation of wave signals $h$ lensed by a BH, we numerically solve Eq.~(\ref{eq:timedomain_wave}) for each $l$-mode in the time-domain using the finite element method (FEM). Since Eq.~(\ref{eq:timedomain_wave}) is second-order in time, each $l$-mode requires two initial conditions: the mode amplitude $u_{l}|_{t=0}$ and its first-order time derivative $\partial_t u_{l }|_{t=0}$. These two conditions can be obtained by decomposing the initial 3D waves $h|_{t=0}$ and $\partial_t h|_{t=0}$ into their corresponding spherical harmonics
\begin{equation}
\begin{aligned}\label{eq:general_ini}
u_{l }(\rho)|_{t=0} &= r(\rho) \int h(\rho, \theta, \phi)|_{t=0} Y^0_l (\theta, \phi)\,\mathrm{d}\Omega\\
\partial_t u_l(\rho) |_{t=0} &= r(\rho) \int \partial_t h (\rho, \theta, \phi) |_{t=0} Y^0_l (\theta, \phi)\,\mathrm{d}\Omega
\end{aligned}\,.
\end{equation}

For certain discretization scheme, providing these two conditions is equivalent to specifying $u_{l}$'s value at the first two time steps. This is because the initial derivative can be approximated using a forward difference format $u_l|_{t=\Delta t} = u_l|_{t=0} + \Delta t \partial_t u_l|_{t=0}$.

The ideal spatial simulation domain for $u_{l}$ needs to span from the event horizon of the lens black hole (BH), located at $\rho / M = 0.5$ in isotropic coordinates, to infinity. In practice, however, simulations can only be performed within a finite domain. To prevent unphysical wave reflections from the domain's edges, we apply absorption boundary conditions at both the inner and outer boundaries. These conditions, also known as non-reflecting or radiation boundary conditions, are expressed as
\begin{equation}\label{eq:abc}
\left( \partial_n u_l + \frac{1}{\tilde{c}} \partial_t u_l \right) \Big|_{\partial \Omega} = 0\,,
\end{equation}
where $\partial \Omega$ is the boundary of the simulation domain and $\partial_n \equiv \hat{n}^i \partial_i$ is the projection of the gradient along the unit normal vector $\hat{n}^i$ at the boundary. However, these absorption boundary conditions work most effectively only in flat spacetime. To maximize their effectiveness, the outer boundary should be positioned sufficiently far from the BH, while the inner boundary needs to be placed close to the event horizon, where the last two terms of Eq.~(\ref{eq:timedomain_wave}) that characterize the dispersion of waves tend to vanish. 

\subsection{Temporal and spatial discretization}
The FEM requires a weak formulation of the problem. We derive this by multiplying Eq.~(\ref{eq:timedomain_wave}) a test function $\varphi$ and then integrating over the simulation domain $\Omega$
\begin{equation}\label{eq:weak}
\begin{aligned}
\langle \varphi, \partial_t^2 u \rangle = \langle \varphi, \tilde{c}^2 \partial_\rho^2 u \rangle + \langle \varphi, \tilde{c}^2 f(u, \partial_\rho u) \rangle\,,
\end{aligned}
\end{equation}
where $\langle u,v \rangle \equiv \int u v\, \mathrm{d}r$ denotes the inner product of the scalar functions $u$ and $v$. Here we omit the lower index $l$ of multipoles for simplicity. The function $f(u, \partial_\rho u)$ represents the last two terms of Eq.~(\ref{eq:timedomain_wave}) divided by $\tilde{c}^2$.

For the first second-order derivative term $\langle \varphi, \tilde{c}^2 \partial_\rho^2 u \rangle$, we apply Gauss's theorem and the absorption boundary condition Eq.~(\ref{eq:abc}). This gives out
\begin{equation}
\begin{aligned}
\langle \varphi, \tilde{c}^2 \partial_\rho^2 u \rangle & = \langle \varphi, \tilde{c}^2\partial_n u \rangle_{\partial\Omega}
 - \langle \partial_\rho (\tilde{c}^2 \varphi), \partial_\rho u \rangle
  \\
&= -\langle \varphi, \tilde{c}\partial_t u \rangle_{\partial\Omega}
 - \langle \partial_\rho (\tilde{c}^2 \varphi), \partial_\rho u \rangle \,.\\
\end{aligned}
\end{equation}
Thus, we can express Eq.~(\ref{eq:weak}) as
\begin{align}
\langle \varphi, \partial_t^2 u \rangle = & -\langle \varphi, \tilde{c}\partial_t u \rangle_{\partial\Omega}
 - \langle \partial_\rho (\tilde{c}^2 \varphi), \partial_\rho u \rangle \nonumber \\
 &+ \langle \varphi, \tilde{c}^2 f(u, \partial_\rho u) \rangle\,. 
\end{align}

For temporal discretization, we adopt the explicit Leapfrog scheme. The time derivatives are discretized as follows
\begin{equation}
\begin{aligned}
\partial_t^2 u^n &= \frac{u^{n+1} - 2u^n + u^{n-1}}{\Delta t^2} \\
\partial_t u^n &= \frac{u^{n+1} - u^{n-1}}{2\Delta t} \\
\end{aligned}\,,
\end{equation}
where $u^{n+1}, u^n, u^{n-1}$ represent the solutions at time steps $t^{n+1}, t^n, t^{n-1}$, respectively. We use a fixed time step length $\Delta t$ throughout the whole evolution. This discretization scheme has second-order accuracy in time and is free from artificial dissipation. 

For spatial discretization, we follow the standard practice of the FEM. The 1D domain $\Omega$ is decomposed into subdomains $\Omega_i$, each consisting of line segments. Each line segment is then divided by several nodes. At each node, we construct a basis function $\varphi_i$ for $i=1,..,N$, where $N$ is the total number of nodes in $\Omega$. Each basis function $\varphi_i$ is non-zero only at its corresponding node and on adjacent line segments (subdomains) and vanishes elsewhere in $\Omega$.  
These basis functions are designed to be orthonomal $\langle \varphi_i, \varphi_j \rangle = \delta_{ij}$ and form a basis for the test function space $\{ \varphi_i \}_{i=0}^N$. 

Given that Eq.~(\ref{eq:weak}) is valid for any test function $\varphi$, we select it to be the basis function $\varphi_i,i=1,..,N$ at each node. This gives out $N$ equations
\begin{align}
&\langle \varphi_i, u^{n+1} \rangle + \frac{\Delta t}{2} \langle \varphi_i, \tilde{c}u^{n+1} \rangle_{\partial\Omega} \nonumber \\
=& \langle \varphi_i, 2u^n - u^{n-1} \rangle + \frac{\Delta t}{2} \langle \varphi_i, \tilde{c}u^{n-1} \rangle_{\partial\Omega} \nonumber \\
-& \Delta t^2 \left[ \langle \tilde{c}^2 \partial_\rho\varphi_i + 2\tilde{c} \partial_\rho \tilde{c} \varphi_i, \partial_\rho u^n \rangle 
- \langle \varphi_i, \tilde{c}^2 f(u^n, \partial_\rho u^n) \rangle \right]\,. \label{eq:spatial_discretization}
\end{align}
Since $\varphi_i$ is only non-zero within a localized region, Eq.~(\ref{eq:weak}) is spatially discretized in this manner.

We further express the solution at each time step as a linear combination of basis functions $u^n = \sum_i U^n_i \varphi_i$, where $U^n_i$ are coefficients of $\varphi_i$ at time $t^n$. Substituting this expansion into Eqs.~(\ref{eq:spatial_discretization}) yields a linear system of equations
\begin{align}
\left( M + \frac{\Delta t}{2} B \right) U^{n+1} &= 
M(2U^n - U^{n-1}) + \frac{\Delta t}{2} B U^{n-1} \nonumber \\
&- \Delta t^2 (A + C - F) U^n\,,
\end{align}
where $M, A, B, C, F$ are $N \times N$ matrices and $U^n$ is the solution vector.
\begin{equation}
\begin{aligned}
M_{ij} & = \langle \varphi_i, \varphi_j \rangle \\
A_{ij} & = \langle \tilde{c}^2\partial_\rho \varphi_i, \partial_\rho \varphi_j \rangle \\
C_{ij} & = \langle 2\tilde{c}\partial_\rho \tilde{c} \varphi_i, \partial_\rho \varphi_j \rangle \\
F_{ij} & = \langle \varphi_i, \tilde{c}^2 f(\varphi_j, \partial_\rho \varphi_j) \rangle \\
B_{ij} & = \langle \varphi_i, \tilde{c}\varphi_j \rangle_{\partial\Omega} \\
U^n &= [U^n_0, U^n_1,..., U^n_N]^T
\end{aligned}\,.
\end{equation}
Given $U^n$ and $U^{n-1}$ at previous two steps,  $U^{n+1}$ can be solved from the linear system.

We implement our simulation code using the C++ FEM library \texttt{deal.II}~\cite{2024:africa.arndt.ea:deal,dealii2019design}. For spatial discretization, we adopt fourth-order Lagrangian elements. By employing the same order Gauss-Lobatto quadrature, the mass matrix $M$ and boundary matrix $B$ become diagonal. This allows for a direct solution of the linear systems, thereby significantly enhancing the efficiency of our time-stepping iterations for the wave equation. To ensure the stability of our explicit time-stepping scheme, we select a time step that is several times smaller than the Courant-Friedrichs-Lewy (CFL) limit.

\subsection{Simulation validation with a hemispherical-shell Gaussian pulse }
In this subsection, we validate our time-domain mode-sum method by comparing its results against those from direct, high-resolution 3D numerical simulations. For this comparison, we simulate the evolution of a left hemispherical shell pulse with a radial Gaussian profile, generated by the following initial conditions
\begin{equation}
\begin{aligned}\label{eq:half_sphere}
\partial_t h|_{t=0} &=
\begin{cases}
\frac{1}{\sqrt{4\pi}r(\rho)}\exp \left[-\frac{(r_*(\rho) - r_{*0})^2}{2\sigma^2} \right]\quad &,\, x < 0 \\
0\quad &,\, x\geq 0
\end{cases}
\\
h|_{t=0}&=0
\end{aligned}\,,
\end{equation}
where $r_*$ is the radius in tortoise coordinate

\begin{align}
    r_* &= r + 2M \log\left(\frac{r}{2M} - 1\right)\nonumber\\
    &=\frac{(M+2\rho)^2}{4\rho}+2M \log\left(\frac{(M+2\rho)^2}{8\rho M} - 1\right)\,.
\label{eq:tortoise}
\end{align}
$r_{*0} = 4M$ and $\sigma = M/2$ are parameters for the Gaussian profile.

\begin{figure*}[htbp]
    \centering
\includegraphics[width=0.95\textwidth]{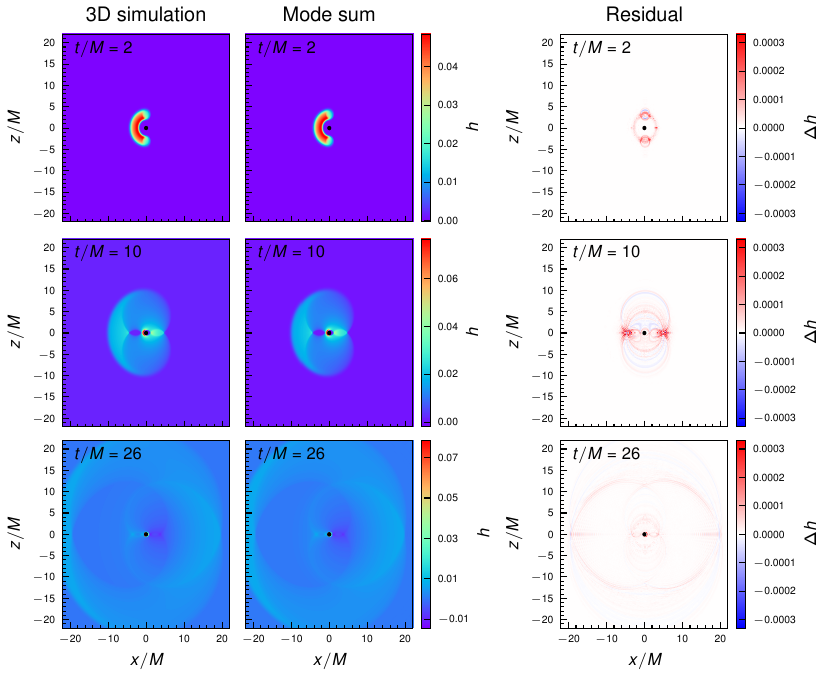} 
\caption{Comparison of a direct 3D numerical simulation and the time-domain mode-sum method. The left column displays snapshots of the wave signals in the $x$-$z$ plane at different times from the direct high-resolution 3D simulation. The middle column shows the corresponding results reconstructed using our time-domain mode-sum method. The right column presents the residuals (the difference) between the two methods. Overall, the time-domain mode-sum method exhibits excellent agreement with the 3D simulation, even in cases of complex wave dynamics. The 3D simulations are in isotropic coordinates of the Schwarzschild BH.
}\label{fig:a1}
\end{figure*}

For the 3D simulation, we solve the wave equation Eq.~(\ref{eq:scalarfield}) using the FEM based on the code developed in our previous work~\cite{He2022PRD}. We discretize the equation using an explicit Leapfrog scheme in time and fourth-order Lagrangian elements in space. We set the simulation domain as a 3D spherical shell extending from an inner radius of $\rho/M=0.52$ to outer radius of $\rho/M=90$. We impose zero-order absorbing inflow and outflow boundary conditions at the inner and outer boundaries respectively ($\mathcal{B}_0$ in \cite{abc}). The simulation has $\sim8.8\times 10^7$ degrees of freedom (DoF). We set the time step to $\Delta t / M = 2 \times 10^{-3}$, which is 60 times smaller than the minimum CFL criteria.
\begin{figure*}[htbp]
    \centering
\includegraphics[width=0.95\textwidth]{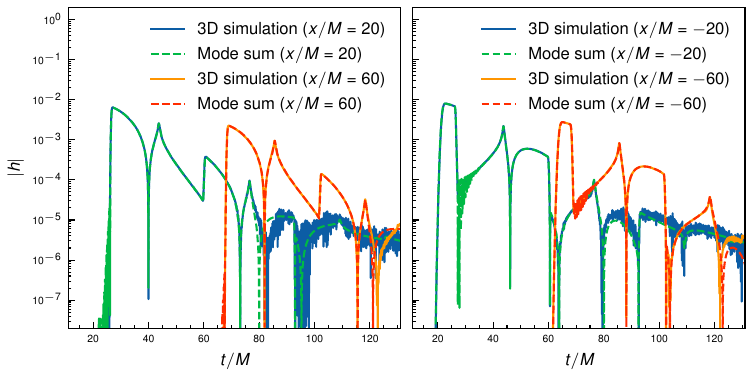} 
\caption{Temporal waveforms observed on the $x$-axis. The left panel shows waveforms in the forward direction at $x/M = 20,\,60$, while the right panel shows waveforms in the backward direction at  $x/M = -20,\,-60$. Waveforms from the direct 3D simulation (solid lines) are compared with those reconstructed using the time-domain mode-sum method (dashed lines) up to $l=63$. The results from the two methods are in excellent agreement.
}\label{fig:a2}
\end{figure*}
For the time-domain mode-sum method, the initial conditions for each $u_l$ are determined by Eq.~(\ref{eq:general_ini}) and Eq.~(\ref{eq:half_sphere}). We use a 1D radial grid that covers the same range as our 3D simulation ( $\rho/M$ from $0.52$ to $90$), but with a finer resolution, resulting in $1 \times 10^4$ DoFs for each $u_l$. We set the time step to $\Delta t / M =2 \times 10^{-3}$, 3 times smaller than the minimum CFL criteria. We compute all the modes up to $l=63$. Since the initial condition at a fixed radius is a hemisphere, only $l=0$ mode and modes with odd numbers of $l$ are non-zero.

\begin{figure*}[htbp]
    \centering
\includegraphics[width=0.90\textwidth]{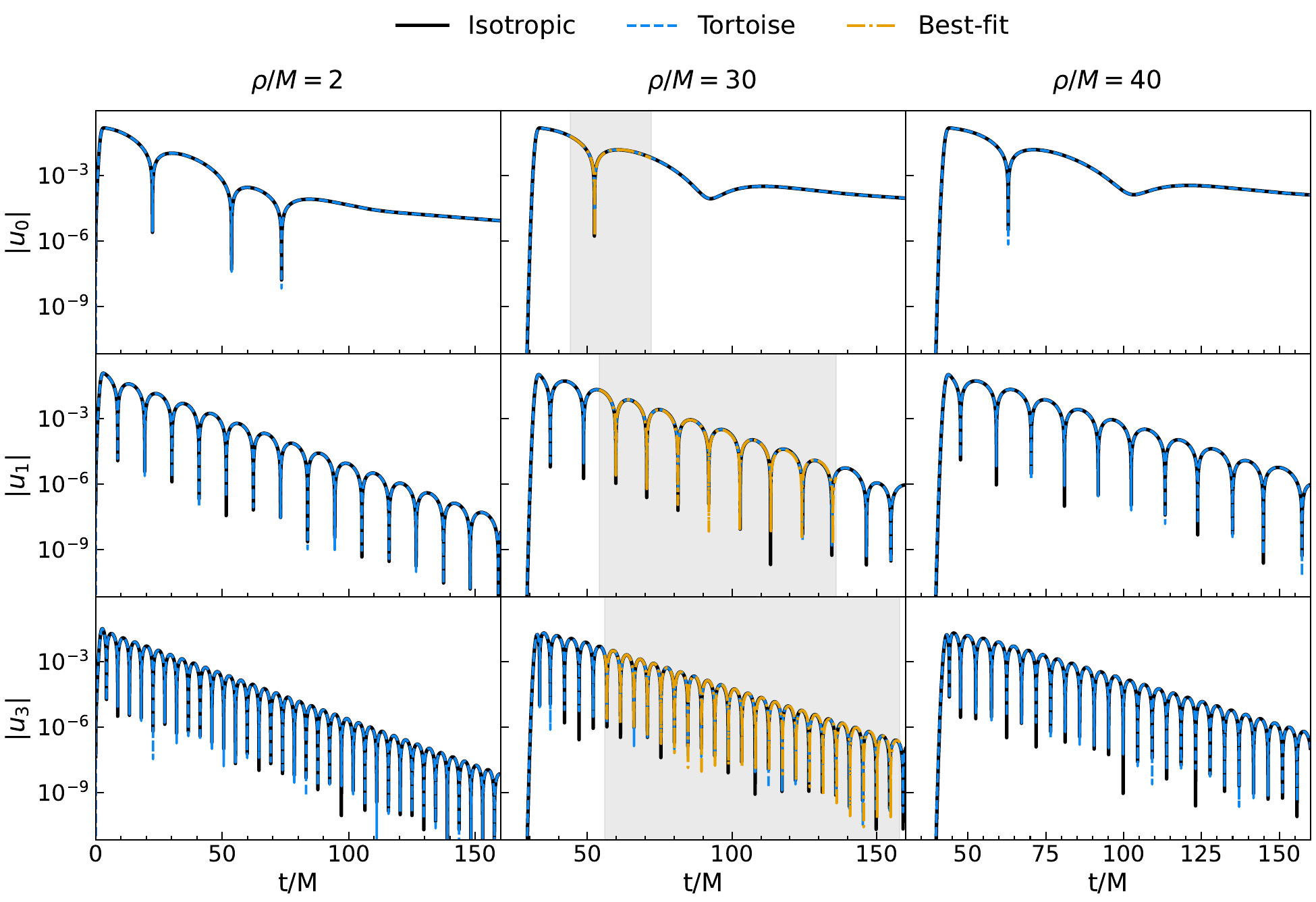} 
    \caption{Comparison of the solutions for $u_l$ between different numerical methods across the first three non-zero modes $l=0,\,1,\,3$. Each row corresponds to a different mode and each column is observed at different radial distances of $\rho/M =2,\, 30, \,40$. The black solid lines show results from our FEM in isotropic coordinates. The blue dashed lines show results from a Finite Difference Method (FDM) in tortoise coordinates with the inner absorbing boundary set at $r_*/M=-40$ (equivalent to $\rho / M \approx 0.50003$). In the middle column, the yellow dash-dotted lines are  the best-fit QNM templates overlaid on $u_l$, fitted over the time intervals indicated by the shaded regions. }
    \label{fig:a3}
\end{figure*}

We present our numerical results using isotropic coordinates, as these coordinates provide a natural representation of the lens BH's geometry in 3D space. Figure~\ref{fig:a1} shows snapshots of the wave signals at different times from a direct 3D simulation (left column) and reconstructed by the time-domain mode-sum method (middle column) respectively. The right column of Figure~\ref{fig:a1} shows the residuals between these two methods. The relative difference is typically below $1\%$. The time-domain mode-sum method exhibits excellent agreement with the 3D simulations, even in cases of complex wave dynamics.

Figure~\ref{fig:a2} shows the temporal waveforms observed by local observers on the $x$-axis. These observers are located at $x/M = 20,\, 60$ and $x/M = -20,\,-60$. The solid lines represent the results from the direct 3D simulation, while the dashed lines correspond to the outcomes from the mode-sum method. Similar to the spatial case, the temporal waveforms show excellent agreement as well.

In addition to the waveforms, we also examine the impact of the absorbing boundaries on our simulations. For the outer boundary, we set it sufficiently far from the lens BH so that signals from the outermost analyzed region could not reach it within the simulation time. This ensures that the outer boundary does not affect our results. To assess the impact of the inner boundary, we compare our FEM results with those derived from an alternative method, in which we solve the wave equation for $u_l$ using the tortoise coordinate with a Finite Difference Method (FDM) based on the Lax-Wendroff scheme, a technique widely adopted in the literature~\cite{Baibhav:2023clw}. In the tortoise coordinate, we set the inner boundary much closer to the event horizon at $r_*/M = -40$ (equivalent to $\rho/M=0.50003$). As shown in Figure~\ref{fig:a3}, the results for $u_l$ from our FEM in isotropic coordinates (solid lines) agree well with those from FDM in tortoise coordinates (dashed lines).

\begin{table}[htbp]
  \centering
  \caption{Best-fit parameters of QNMs\label{QNM_results_half}}
  \begin{tabular}{c|c|c}
    \hline
    mode & $M\omega_{R}+iM\omega_{I}$ (Theory\cite{Mamani:2022akq,wardell_2025_17194969}) & $M\omega_{R} + iM\omega_{I} $(best-fit) \\
    \hline
    $l=0$ & $0.110455-0.104896i$ & $0.110665- 0.084627i$ \\
    \hline
    $l=1$ & $0.292936-0.097660i$ & $0.293011-0.098518i$ \\
    \hline
    $l=3$ & $0.675366-0.096500i$ & $0.672454-0.096325i$ \\
    \hline
    $l=5$ & $1.05961-0.096337i$ & $1.05951-0.096336i$ \\
    \hline
    $l=11$ & $2.21372-0.096251i$ & $2.21385-0.096228i$ \\
    \hline
    $l=21$ & $4.13797-0.096232i$ & $4.13787-0.096286i$ \\
    \hline
  \end{tabular}
\end{table}

The middle column of Figure~\ref{fig:a3} shows the best-fit QNM templates (yellow dash-dotted lines) overlaid on $u_l$ at $\rho/M= 30$. The numerical results of $u_l$ are fitted using a single frequency damped sinusoid model~\cite{Baibhav:2023clw}. The shaded regions indicate the time interval used for fitting. Table~\ref{QNM_results_half} compares the best-fit parameters with theoretical predictions (fundamental modes without overtones)~\cite{Mamani:2022akq,wardell_2025_17194969}. In the table, we include more numerical results up to $l=21$. The numerical solutions for $u_l$ agree with the theoretical predictions of QNM very well. This confirms that the late-time behavior of $u_l$ is QNMs, which in turn further validates the robustness of our numerical method.

\section{3D waveform of BH lensing} After testing our numerical method, we perform the numerical simulation of a scalar wave lensed by a Schwarzschild BH. We assume that the source of the incident wave is located far from the lens BH at a distance greater than a kiloparsec. When analyzing the paraxial waves near the optical axis, this large separation makes it reasonable to approximate the incoming wave as a plane wave. To investigate the excitation of QNMs in BH lensing, we consider the incident signal as a single burst with Hanning-profile, given by $\sin^2(\pi x/X)$. The signal width is set to $X=4M$. For a lens BH with a mass of 100 $M_\odot$, this width corresponds to a signal duration of approximately $\sim$2 ms, a timescale typical for astrophysical events such as supernova explosions or the final merger stage of binary neutron stars. In our simulations, the lens BH is positioned at the origin, while the initial incident wave packet starts at $x/M = -80$ and propagates along the $x$-axis. As demonstrated in Appendix, this distance is sufficiently large to ensure that the gravitational influence of the lens BH on the initial incident wave is negligible. For the mode-sum method, we calculate the evolution of all modes $u_l$ up to $l = 127$. A detailed analysis of the convergence of our mode-sum method is provided in Appendix.

\begin{figure*}[htbp]
    \centering
\includegraphics[width=0.95\textwidth]{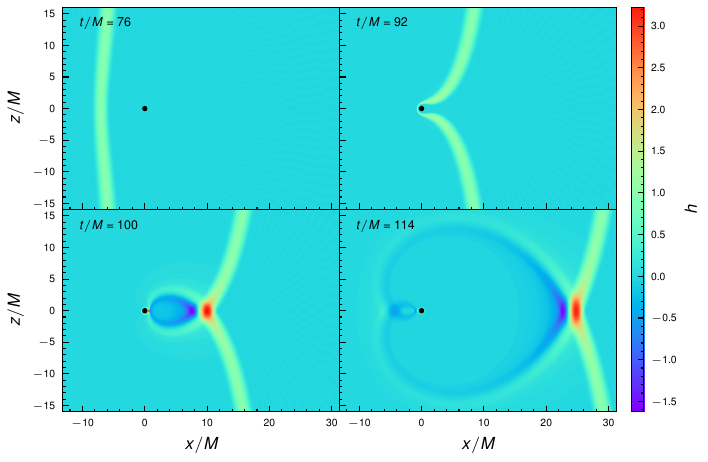} 
\caption{Snapshots of the 3D numerical results in the $x$-$z$ plane obtained from the time-domain mode-sum method. The solid black disk represents the location of the BH, with its radius indicating the event horizon. The color bar to the right shows the amplitude of these wave signals. Each panel illustrates the lensed wave signals at different times. The 3D simulations are in isotropic coordinates of the Schwarzschild BH.}
    \label{fig:1}
\end{figure*}

Figure~\ref{fig:1} shows snapshots of our numerical results in the $x$-$z$ plane obtained from mode-sum method at different times, where wave signals are presented using isotropic coordinates of the Schwarzschild BH. The black disk denotes the location of the lens BH, and its radius represents the event horizon. The color bar indicates the amplitude of the scalar wave. As shown in~\cite{He2022PRD}, the leading wavefront of the incident wavelet is a null hypersurface that propagates along null geodesics. The trajectories of these geodesics are bent by the gravitational field of the black hole. Consequently, after the incoming scalar wave interacts with the BH (for $t/M>92$), its wavefront becomes significantly distorted. This distortion leads to self-intersection and the formation of caustics. This phenomenon closely resembles the complex structure of the retarded Green's function in the geometric optics limit, as investigated in~\cite{PhysRevD.84.104002}, which characterizes wave propagation between two points in a black hole spacetime. Notably, a strong wavepacket forms along the $x$-axis in the forward direction. Furthermore, as shown in the lower-right panel of Fig.~\ref{fig:1} at a later time ($t/M>114$), waves near the photon sphere are observed encircling the BH. These orbiting waves subsequently generate a secondary signal that propagates back in the opposite direction.

\begin{figure}[H]
    \centering
\includegraphics[width=0.9\linewidth]{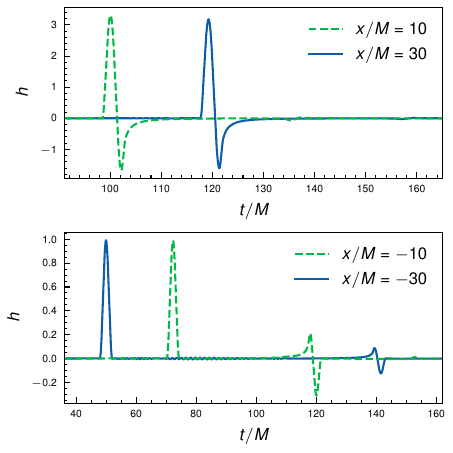} 
\caption{Temporal waveforms observed on the $x$-axis at $x/M=10,\,30$ in forward direction (upper-panel) and $x/M=-10,\,-30$ in backward direction (lower-panel).
}\label{fig:2}
\end{figure}

Figure~\ref{fig:2} shows the temporal waveforms $h$ observed on the $x$-axis. The observers are located at $x/M = 10,\,30$ in the forward direction (upper-panel), and at $x/M = -10,\,-30$ in the backward direction (lower-panel). In the forward direction, the BH significantly distorts the temporal waveform compared to the original incident waveform, forming a tail-like trailing signal without obvious damped oscillations. In the backward direction, the primary signal is simply the incident wave approaching the BH, followed by a secondary signal that arises from waves that have circled around the BH and then traveled back to the observer (see the lower-right panel in Figure~\ref{fig:1}). 

\section{QNM analysis in lensed waveform} 

\begin{figure*}[htbp]
    \centering
\includegraphics[width=0.9\textwidth]{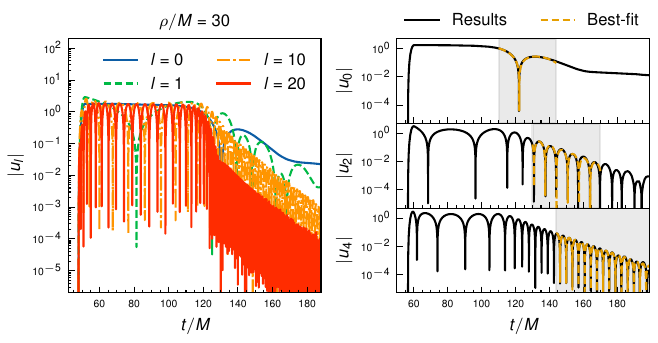} 
\caption{Left: The time evolution of  $u_l(t,\rho)$ at $\rho/M=30 $ for $l=0,1,10,20\,$.The amplitude of the $l=10$ mode is approximately $10\%$ that of the $l=0$ mode, and even for $l=20$, it still remains at a significant percentage level relative to the $l=0$ mode. Right: Comparison of the best-fit QNM templates (yellow dashed lines) with $u_l(t)$ (solid black lines) for $l=0,2,4$. The shaded regions indicate the time interval used for fitting. }
    \label{fig:3}
\end{figure*}

A key advantage of our method is that the temporal waveforms shown in Figure~\ref{fig:2} are superposition of multipoles $u_l(t,\rho)$ according to Eq.~(\ref{eq:Spherical_Harmonic}). This allows us to directly perform QNM mode analysis on these $u_l(t,\rho)$ multipoles. In contrast, conventional analysis in numerical relativity first compute the total gravitational wave signals and then extract the individual multipole components of waveforms (see~\cite{Bishop:2016lgv} for reviews). This could add extra noise depending on the method used.

The left panel of Figure~\ref{fig:3} shows the time evolution of  $u_l(t,\rho)$ at $\rho/M=30$ across different $l$ modes. A notable feature of BH lensing is the excitation of a substantial number of high-$l$ modes. For instance, the amplitude of the $l=10$ mode is approximately $10\%$ that of the $l=0$ mode, and even for $l=20$, it remains at a significant percentage level relative to the $l=0$ mode. 

\begin{table}[htbp]
  \centering
  \caption{Best-fit QNM parameters \label{QNM_results}}
  \begin{tabular}{c|c|c}
    \hline
    mode & $M\omega_{R}+iM\omega_{I}$ (Theory\cite{Mamani:2022akq}) & $M\omega_{R} + iM\omega_{I} $(Best-fit) \\
    \hline
    $l=0$ & $0.110455-0.104896i$ & $0.105459-0.051805i$ \\
    \hline
    $l=2$ & $0.483644-0.096759i$ & $0.474154-0.096738i$ \\
    \hline
    $l=4$ & $0.867416-0.096392i$ & $0.864519-0.096839i$ \\
    \hline
    $l=6$ & $1.25189-0.096305i$ & $1.25194-0.096237i$ \\
    \hline
    $l=10$ & $2.02132-0.096256i$ & $2.02108-0.096426i$ \\
    \hline
    $l=20$ & $3.94553-0.096233i$ & $3.94575-0.094824i$ \\
    \hline
  \end{tabular}
\end{table}

Unlike the conventional time-domain QNM paradigm, $u_l$ for the lensed waves typically exhibits a significantly longer prompt signal. To verify that the observed late-time ``rings" are indeed QNMs, we fit $u_l(t,\rho)$ using a single frequency damped sinusoid model~\cite{Baibhav:2023clw}. The right panel of Figure~\ref{fig:3} shows these best-fit QNM templates (yellow dashed lines) overlaid on  $u_l(t)$ at $\rho/M= 30$ for $l=0,2,4$. The shaded regions indicate the time interval used for fitting. Table~\ref{QNM_results} compares the best-fit parameters with theoretical predictions (fundamental modes without overtones)~\cite{Mamani:2022akq,wardell_2025_17194969}.
The overall QNM frequencies ($\omega_R$) show well agreement with theory while the decay rate ($\omega_I$) for the $l=0$ mode exhibits some discrepancy. This inconsistency arises because, for a scalar field, the late-time tail has a significant impact especially on low $l$ mode~\cite{PhysRevD.5.2419}. 

\begin{figure}[H]
    \centering
\includegraphics[width=0.99\linewidth]{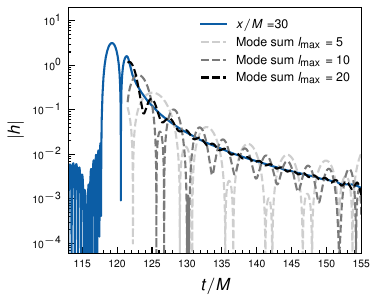} 
    \caption{The absolute amplitude of the lensed temporal waveform $|h|$ observed at $x/M=30$ on the $x$-axis (solid blue line). The dashed lines show the waveforms reconstructed by truncated mode sum of $u_l$ up to $l_{\rm max}=5,\,10,\,20$ respectively (from lighter to darker). Because the individual oscillations in each mode largely cancel out, the final reconstructed waveform does not exhibit oscillatory feature during the QNM stage (e.g. $120 \lesssim t/M \lesssim 160$).}
    \label{fig:41}
\end{figure}

Figure~\ref{fig:41} shows the reconstructed lensed temporal waveforms observed at $x/M=30$ on the $x$-axis (similar to Figure~\ref{fig:2}, but for the absolute value $|h|$). Unlike typical QNM signals that are characterized by damped oscillations (illustrated in Figure~\ref{fig:3} for various multipoles $u_l$), the reconstructed waveforms do not exhibit such oscillations during the QNM stage (e.g. $120 \lesssim t/M \lesssim 160$). This occurs because the overall waveform is a superposition of multiple QNMs, which results in oscillations that largely cancel each other out. To illustrate this, we reconstruct the waveforms by summing modes $u_l$ up to $l_{\rm max}=5,\,10,\,20$ (represented by progressively darker dashed lines). As more modes are included (i.e. as $l_{\rm max}$ increases), the oscillations diminish.

\section{Gaussian beam} 
\begin{figure*}[htbp]
    \centering
\includegraphics[width=0.9\textwidth]{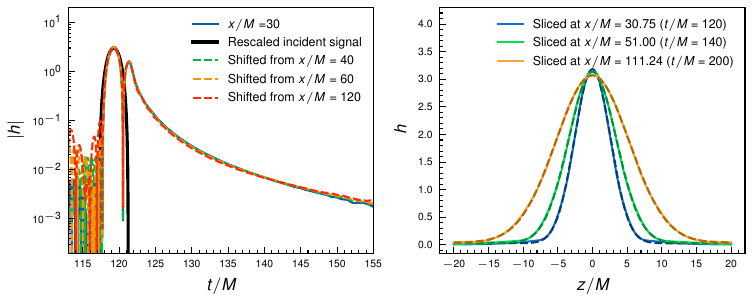} 
\caption{Lensed waveforms and paraxial Gaussian beam. Left: The absolute amplitude of the lensed waveforms $|h|$ as observed at different radii on the $x$-axis. For comparison, all waveforms are shifted in time to align their peaks with the peak of the waveform at $x/M=30$. The lensed waveforms remain nearly unchanged at the different radii. The solid black line shows the incident waveform with amplitude rescaled to match the lensed waveform. Right: Cross-sectional intensity profiles of the paraxial lensed beam along the $z$-axis (solid lines). Each profile is captured at the time of the beam's peak intensity.  The dashed lines show the best-fit Gaussian models, which precisely match the simulated profiles.}\label{fig:5}
\end{figure*}

To investigate how the lensed waveform's amplitude scales with the distance from the BH, Figure~\ref{fig:5} compares waveforms at several positions on the $x$-axis ($x/M=30,\,40,\,60,\,120$). The waveforms are time-shifted to align their peaks with that of the waveform at $x/M=30$. We observe that the amplitude does not decay with distance as a spherical wave would. Instead, the lensed waveforms in the simulated near-field region exhibit nearly constant amplitude and shape.

This behavior indicates that interference among the lensed paraxial waves forms a Gaussian beam. To demonstrate this, the right panel of Figure~\ref{fig:5} shows the beam's spatial intensity profiles (solid lines) measured along the transverse $z$-axis at three beam peak positions at different times. The dashed lines show the best-fit Gaussian models, which precisely match the simulated profiles. This demonstrates that the lensed paraxial waves form a coherent Gaussian beam (e.g.~\cite{Born_Wolf_2019} for a detailed discussion of Gaussian beam properties).

\section{Discussions} In this work, we investigate the excitation of QNMs in gravitational lensing of a BH using a scalar field model. We simulate an incident burst signal lensed by a Schwarzschild BH via the time-domain mode-sum method in 3D spacetime. We find that the burst signal excites a substantial number of high-$l$ modes. Even for $l=20$, the amplitude of the mode remains at a percentage level relative to the $l=0$ mode. We confirm that the late-time behaviors of these modes are indeed QNMs by fitting the simulation data to QNM templates.

This QNM excitation is fundamentally different from the resonant excitation discussed in the literature~\cite{PhysRevD.110.124011,WOS:A1972L790000003,PhysRevD.79.064016,PhysRevD.30.295,PhysRevD.31.290}, which is caused by the interference between incoming waves and those that wind around the photon sphere. This resonant excitation is most prominent for high-$l$ modes in the eikonal (ray optics) limit, where QNMs are interpreted as null particles trapped in unstable orbits at the photon sphere that gradually leak out. In contrast, the QNMs we observe do not arise from such interference. As illustrated in Fig.~\ref{fig:1}, the primary lensed signals are generated by waves passing far from the photon sphere. Due to the significant time delay, the orbiting waves trapped around the photon sphere lag far behind the primary signal, preventing them from interfering with the primary signal.

Indeed, within the theoretical framework of QNM based on Laplace analysis(e.g.~\cite{Hans-Peter_Nollert_1999,Kokkotas:1999bd,Berti_2009,RevModPhys.83.793, Berti:2025hly}), the QNM Green’s function remains non-zero regardless of the initial conditions, which implies that a direct excitation process is at play. Our observation of QNMs across a broad spectrum of $l$-modes aligns well with theoretical expectations. We refer to this direct excitation mechanism as “non-resonant excitation”, which is effective even for low-$l$ mode, including $l=0$ mode.

Moreover, we find that the superposition of a large number of multipole QNMs leads to oscillations that largely cancel each other out. As a result, the lensed temporal waveforms do not exhibit the characteristic oscillatory patterns typically anticipated.

In addition to exciting QNMs, we find that the lensed waves form a highly directional and coherent Gaussian beam along the $x$-axis. This beam’s cross-sectional intensity follows a precise Gaussian profile. Our simulations show that the lensed temporal waveforms and amplitudes remain nearly unchanged across the near-field region. 
Based on Gaussian beam properties, we expect the cross-sectional intensity to stay Gaussian as the beam propagates into the far-field. However, the beam will spread due to diffraction, and the wavefronts will become planar. Despite the complex changes in the beam's amplitude, the shape of the temporal waveform remains preserved. This preservation enables a distant observer to detect the distinctive features of the lensed temporal waveform, making these signals potentially identifiable. Such observations could open a new path for testing the linear perturbation theory of BHs.

Finally, we argue that although this work investigates the QNMs in gravitational lensing of BH using a scalar field model, we expect similar qualitative results for gravitational wave lensing. This is due to the similarity between the scalar field equation Eq.~(\ref{eq:Spherical_Harmonic}) and Regge-Wheeler~\cite{Regge_Wheeler} or Zerilli-Moncrief~\cite{Zerillia,Zerillib,MONCRIEF1974343} equations. A detailed analysis on tensorial gravitational waves will be presented in our follow-up papers. 

\vspace{5mm}
\textbf{Acknowledgments}
This work is supported by the National Key R$\&$D Program of China (Grant No. 2021YFC2203002), the National Natural Science Foundation of China (Grants No. 12475058, No. 124B2093). This work makes use of the Black Hole Perturbation Toolkit.

\section*{Appendix}
In this appendix, we perform the convergence tests of the numerical results on BH lensing. We assume that the wave source is located far from the lens BH($>{\rm kpc}$). When restricting our analysis to paraxial waves and examining signals within a few hundred seconds of the optical axis, we can reasonably approximate the incoming wave signal as a plane wave. However, due to the long-range nature of gravity, the gravity of the lens BH may distort the incident wavefront and invalidate this approximation. Therefore, it is important to test that the initial wave is set far enough from the lens BH so that it will not significantly alter the simulation results.

To perform such a test, similar to what is described in the main text, we assume that the incident wave initially propagates in flat spacetime along the $x$-axis with a Hanning waveform
\begin{equation}\label{eq:plane_hanning}
h 
=
\begin{cases}
\sin^2 \left[\frac{\pi (x - x_0- \tilde{c}t)}{X} \right]\,&, \, x \in [x_0 +\tilde{c}t, X +x_0+ \tilde{c}t] \\
 0\, &, \, \mathrm{Otherwise}
\end{cases}\,,
\end{equation}
where $x_0$ is the initial position. To apply the mode-sum method, we set the initial conditions for the wave equation of $u_l$ using Eq.~(\ref{eq:general_ini}) and Eq.~(\ref{eq:plane_hanning}). We use a one-dimensional mesh extending from $\rho/M = 0.51$ to $\rho / M = 440$. The finite element discretization of this mesh yields $1.6\times 10^4$ DoFs. We set the time step to $\Delta t/M = 2 \times 10^{-3}$, which is 4 times smaller than the minimum CFL criteria. We compute all the modes up to $l=127$.

\begin{figure*}[htbp]
    \centering
\includegraphics[width=0.95\textwidth]{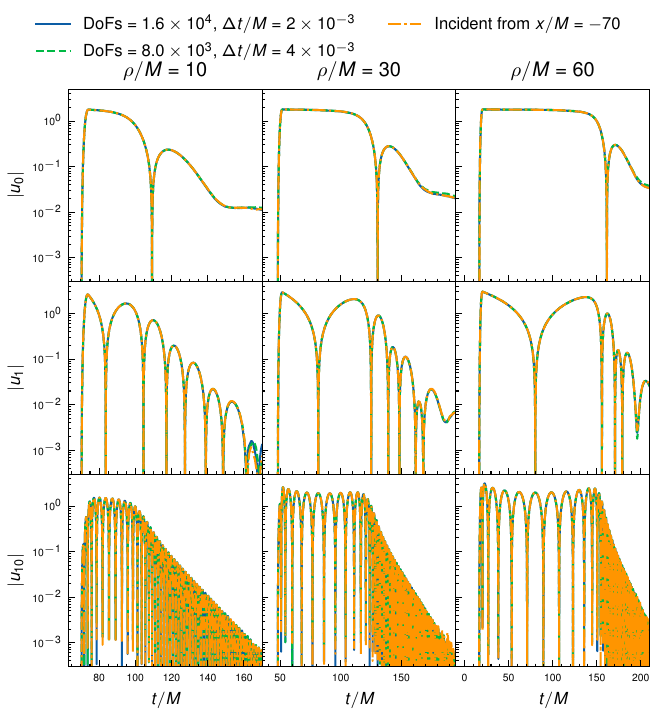} 
    \caption{Solutions of $u_l$ for three different modes $l=0,\,1,\,10$ (each row respectively) at three different radial distances $\rho/M=10,\,30,\,60$ (each column respectively). The blue solid lines represent simulation results with the initial incident wave starting at $x/M=-80$, while the yellow dash-dotted lines represent the incident wave initiated at $x/M=-70$, whose waveforms are also shifted in time for comparison. The close match between these two scenarios confirms that starting the incident wave at $x/M=-80$, as used in the main text, is sufficiently far from the lens BH. The green dashed lines show results obtained by halving the spatial and temporal resolution. The agreement between the full and reduced resolutions validates that the resolution used in our analysis in the main text is adequate.}
    \label{fig:a5}
\end{figure*}

Figure~\ref{fig:a5} shows solutions of $u_l$ across modes $l=0,\,1,\,10$ extracted at $\rho/M=10,\,30,\,60$, respectively. The blue solid lines represent the case where $x_0/M=-80$, while the yellow dash-dotted lines correspond to $x_0/M=-70$. The agreement between these two cases demonstrates that placing the initial incident wave at a greater distance has negligible impact on the results, which confirms that $x_0/M=-80$, as adopted in the main text, is sufficiently distant from the lens BH.

To further test the numerical convergence of our simulations. We perform simulations by halving the spatial and temporal resolution. The results (green dashed lines in Figure~\ref{fig:a5} show great consistency compared to the original resolution. The relative difference is $\sim 10^{-4}$ when $|u_l| > 10^{-3}$, which is adequate for our analysis in the main text.

\begin{figure*}[htbp]
    \centering
\includegraphics[width=0.95\textwidth]{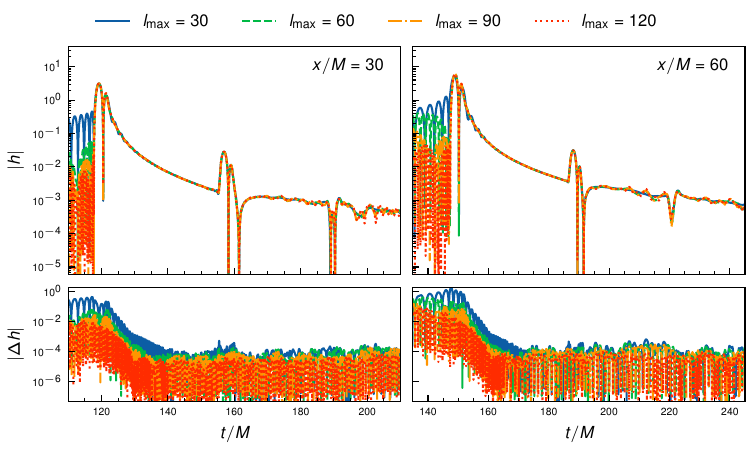}  
    \caption{Upper panels: the reconstructed temporal waveforms $h$ at two observer locations on the $x$-axis, $x/M=30$ and $x/M= 60$. Different line styles represent waveforms mode-summed up to $l_{\rm max} = 30,\,60,\,90,\,120$. Lower panels: the corresponding variations relative to the waveform computed with $l_\mathrm{max}=127$. The waveforms converge as $l_{\rm max}$ is increased. For $l_{\rm max}\geq 120$, typical relative variations $|\Delta h / h|$ are below $10^{-2}$ across most of the waveform's duration (\textit{e.g.} $120 < t/M < 160$ in the left panel).}
    \label{fig:a6}
\end{figure*}

Next, we test the convergence of the mode-sum for the temporal waveform $h$. As defined in Eq.~(\ref{eq:Spherical_Harmonic}), this waveform is reconstructed by summing the time-evolved functions $u_l(t,\rho)$, each weighted by its corresponding spherical harmonics. The upper panels of Figure~\ref{fig:a6} show the reconstructed temporal waveforms $h$ observed on the $x$-axis at $x/M=30$ and $x/M= 60$, respectively. The different line styles represent waveforms mode-summed up to  $l_{\rm max}=30,\,60,\,90,\,120$. The lower panels show the corresponding variation relative to the waveform computed with $l_\mathrm{max} = 127$. As $l_{\rm max}$ increased, the waveforms gradually converge. For $l_{\rm max}\geq 120$, typical relative variations $|\Delta h / h|$ are below $10^{-2}$ across most of the waveform's duration (\textit{e.g.} $120 < t/M < 160$ in the left panel).

\bibliography{bibliography}

\end{document}